%% file: sample-sigconf.tex
\documentclass[sigconf]{acmart}

\usepackage{booktabs} 
\usepackage{algorithm,algorithmic}
\usepackage{multirow,multicol}

\setcopyright{rightsretained}

\acmDOI{10.475/123_4}

\acmISBN{123-4567-24-567/08/06}

\acmConference[Conference'19]{ACM conference}{2019}{ACM conference}
\acmYear{1997}
\copyrightyear{2016}

\acmArticle{4}
\acmPrice{15.00}

\editor{Jennifer B. Sartor}
\editor{Theo D'Hondt}
\editor{Wolfgang De Meuter}

\begin{document}
\title{Joint Cluster Unary Loss for Efficient Cross-Modal Hashing}

\author{Shifeng Zhang, Jianmin Li and Bo Zhang}
\affiliation{%
  \institution{Institute for Artificial Intelligence, State Key Lab of Intelligent Technology and Systems, \\
  Beijing National Research Center for Information Science and Technology, \\
  Department of Computer Science and Technology, Tsinghua University}
  \streetaddress{}
  \city{Beijing}
  \state{China}
  \postcode{100084}
}
\email{zhangsf15@mails.tsinghua.edu.cn, lijianmin@mail.tsinghua.edu.cn, dcszb@mail.tsinghua.edu.cn}

\renewcommand{\shortauthors}{Zhang et. al.}

\begin{abstract}
With the rapid growth of various types of multimodal data, cross-modal deep hashing has received broad attention for solving cross-modal retrieval problems efficiently. Most cross-modal hashing methods follow the traditional supervised hashing framework in which the $O(n^2)$ data pairs and $O(n^3)$ data triplets are generated for training, but the training procedure is less efficient because the complexity is high for large-scale dataset. To address these issues, we propose a novel and efficient cross-modal hashing algorithm in which the unary loss is introduced. First of all, We introduce the Cross-Modal Unary Loss (CMUL) with $O(n)$ complexity to bridge the traditional triplet loss and classification-based unary loss. A more accurate bound of the triplet loss for structured multilabel data is also proposed in CMUL. Second, we propose the novel Joint Cluster Cross-Modal Hashing (JCCH) algorithm for efficient hash learning, in which the CMUL is involved. The resultant hashcodes form several clusters in which the hashcodes in the same cluster share similar semantic information, and the heterogeneity gap on different modalities is diminished by sharing the clusters. The proposed algorithm is able to be applied to various types of data, and experiments on large-scale datasets show that the proposed method is superior over or comparable with state-of-the-art cross-modal hashing methods, and training with the proposed method is more efficient than others.
\end{abstract}

%
%



\maketitle

\input{samplebody-conf}

\bibliographystyle{ACM-Reference-Format}
\bibliography{sample-bibliography}

\end{document}

%% file: samplebody-conf.tex
\section{Introduction}

With the rapid growth of various types of multimedia data, Approximate Nearest Neighbor (ANN) search~\cite{gionis1999similarity} has received broad attention for fast information retrieval. Hashing is a popular tool for ANN search in which the data are encoded into compact short binary codes, and it can perform efficient retrieval due to its fast distance computation and small storage space~\cite{gionis1999similarity,datar2004locality,ji2013min,liu2012supervised,xia2014supervised,liu2016deep}. 

In practical applications, there may be various modalities among the data. For example, the images may have tags or image descriptions, a document has multiple languages, etc. Cross-modal retrieval is a fundamental problem in the real world. Given the query of a certain modality, we should search for relevant data of another modality. Cross-modal hashing is a good approach for solving the retrieval problems~\cite{kumar2011learning,lin2015semantics,ding2014collective,zhang2014large}. By encoding the data of different modalities to short binary codes, we can perform the cross-modal retrieval efficiently by searching the hashcodes. Among the existing cross-modal hashing methods, deep cross-modal hashing have witnessed great success~\cite{jiang2017deep,zhang2018attention,cao2016correlation,zhao2017tuch,li2018self,liu2017deep,zhang2018generative,cao2016visual,liong2017cross}. The hashcodes of different modalities are encoded with deep neural networks, thus the genereted codes contain more semantic information to achieve better retrieval results.

For (deep) cross-modal hashing methods, the basic problem is to reduce the heterogeneity gap between different modalities. An efficient and popular way is to model the hamming distance between the hashcode of one modality and that of another modality. DCMH~\cite{jiang2017deep}, CHN~\cite{cao2016correlation}, PRDH~\cite{yang2017pairwise} and CMHH~\cite{cao2018cross} propose the pairwise loss for training the hashcodes, ADAH~\cite{zhang2018attention} proposes triplet loss for training. These methods aim at minimizing the hamming distances between hashcodes of different modalities for similar data pairs, and vice versa. Recently, some algorithms propose generation-based methods to align different modalities. TUCH~\cite{zhao2017tuch} and GDH~\cite{zhang2018generative} turns cross-modal hashing to single-modal hashing, in which a text-to-image synthesis model with Generative Adversarial Network (GAN)~\cite{goodfellow2014generative} is introduced. ACMR~\cite{wang2017adversarial}, CYC-DGH~\cite{wu2019cycle} and SSAH~\cite{li2018self} introduce GAN to align the distribution of codes in different modalities. Reducing the heterogeneity gap is able to improve the performance of cross-modal retrieval. On the other hand, it is clear that the pairwise/triplet losses play the crucial role in existing cross-modal hashing algorithms, which is similar with the single-modal supervised hashing algorithms~\cite{lai2015simultaneous,zhu2016deep,cao2016deep}. Therefore, these methods involve at least $O(n^2)$ data pairs and $O(n^3)$ data triplets, making the training procedure less efficient~\cite{zhang2018semantic,zhuang2016fast}.

To overcome the above issue, in (deep) supervised hashing problems, for dataset with semantic labels or tags, some algorithms propose $O(n)$ algorithms in which the labels are regarded as the supervision. CNNBH~\cite{shen2015supervised} directly regards the activation of intermediate layer of a classification model as the hashcodes, but it lack the theoretical relationship between the classifiers and the hamming space. SCDH~\cite{zhang2018semantic} proposes the Unary Upper Bound to bridge the theoretical relationship between the triplet loss and the unary classification-based loss and achieves impressive retrieval results. The derivation strategy of SCDH may be applied to cross-modal hashing algorithm to reduce the training complexity. However, SCDH lies in the assumption that the semantic labels are evenly distributed and different labels are uncorrelated, especially in the multilabel dataset. But in real applications, there are large number of multilabel dataset, and the labels may be highly unbiased and correlated, for example, an image with label "hat" is likely to have label "person". We name it as "{\em structured multilabel data}", and the Unary Upper Bound in SCDH is not accurate for these dataset. We need to explore better Unary Upper Bound for learning with structured multilabel data, so as the cross-modal hashing problems. For cross-modal hashing, some algorithms like SSAH and ACMR introduce label information for training, but the pairwise losses are regarded as the key loss for learning. Motivated by the unary losses used in supervised hashing, efficient cross-modal hashing algorithm with just $O(n)$ training complexity is expected to be discovered.

\subsection{Our Proposal}

In this paper, we propose a novel and efficient cross-modal hashing algorithm with $O(n)$ training complexity. First, we examine the limitations of SCDH~\cite{zhang2018semantic}, and therefore provide an improved Unary Upper Bound (UUB) for structured multilabel data in which the labels are highly unbalanced and correlated. The improved UUB is a more general and accurate bound of the triplet ranking loss. Second, based on the improved UUB, we introduce the Cross-Modal Unary Loss (CMUL), establishing the theoretical relationship between the cross-modal triplet loss~\cite{zhang2018attention} and the $O(n)$ classification-based unary loss. The CMUL conveys that the model of each modality can be trained separately, except that the hashcodes of different modalities are aligned by sharing the auxiliary semantic cluster centers where similar data correspond to the same center. Third, based on CMUL, we propose a novel cross-modal hashing algorithm called Joint Cluster Cross-Modal Hashing (JCCH). The complexity of JCCH is just $O(n)$ and the training procedure is expected to be efficient.

Our main contributions are summarized as follows:

\begin{enumerate}
    \item We introduce the improved Unary Upper Bound of triplet loss for structured multilabel data. This bound is more general and accurate in real applications in which the semantic labels are unbalanced and correlated.
    \item We propose the efficient Cross-modal Unary Loss (CMUL) to bridge the cross-modal triplet loss and classification-based unary loss. We introduce the novel Joint Cluster Cross-modal Hashing (JCCH) based on CMUL. The complexity of CMUL and JCCH is $O(n)$, which can be trained efficiently.
    \item Experiments on large-scale datasets show that the proposed method is superior over or comparable with state-of-the-art cross-modal hashing methods, and the training is more efficient than others.
\end{enumerate}

\section{Cross-modal Unary Loss}

Suppose we are given $n$ data samples, and each sample corresponds to the two modalities $\mathbf{X}_i: (\mathbf{x}_{i1}, \mathbf{x}_{i2}), i=1,2,...,n$. The goal of hash learning is to learn the hash functions $H_1: \mathbf{x}_{i1} \to \{-1,1\}^r, H_2: \mathbf{x}_{i2} \to \{-1,1\}^r$. Denote $Y_i \subseteq \{ 1,...,C \}$ as the labels of data sample $\mathbf{X}_i$, where $C$ is the number of semantic labels. For multiclass dataset, $|Y_i|=1$. The data pairs are similar if they share at least one semantic label. 

In this paper, we want to solve the cross-modal retrieval problem. For two modalities, we aim at searching the relevant data of one modality given the query of the other modality. Formally, we want to retrieve from $\mathbf{x}_{11}, ..., \mathbf{x}_{n1}$ given the query $\mathbf{x_{i2}}$, or vice versa. 

Cross-modal hashing is a popular tool for solving the cross-modal retrieval problem~\cite{kumar2011learning,jiang2017deep}. The goal of hash learning is to learn the hash function of each modality. Denote $F_1(\cdot)$ as the learned function for the first modality and $F_2(\cdot)$ for the second modality, the generated codes are $\mathbf{h}_{i1} = \mathrm{sgn}(F_1(\mathbf{x}_{i1})), \mathbf{h}_{i2} = \mathrm{sgn}(F_2(\mathbf{x}_{i2}))$ respectively. The retrieval process can be performed by computing the hamming distances of the hashcodes from different modalities. 

Inspired by~\cite{zhang2018semantic}, we find that similar derivations can be applied for computing the bound of cross-modal triplet loss~\cite{zhang2018attention}. In this section, we first of all review the single modal hashing problem. We revisit the derivations of UUB in Sec. \ref{sec:uub_recap} and find that the UUB in~\cite{zhang2018semantic} is not accurate for the structured multilabel data in which the labels are unbalanced and correlated. The improved UUB is introduced in Sec. \ref{sec:improve_uub}. For single modality, we denote $\mathbf{x}_{i} \doteq \mathbf{x}_{i1}$ and $F(\cdot) \doteq F_1(\cdot)$ for simplicity, thus the generated codes are $\mathbf{h}_i =  \mathrm{sgn}(F(\mathbf{x}_i)), i=1,2,...,n$. Based on the improved UUB, we propose the UUB for cross-modal triplet loss for cross-modal hashing in Sec. \ref{sec:uub_cross}, which is named as Cross-modal Unary Loss (CMUL).

\subsection{The Unary Upper Bound Recap}
\label{sec:uub_recap}

Consider hashing on single modal dataset, a widely used loss for hash learning is triplet ranking loss~\cite{norouzi2012hamming,lai2015simultaneous}. Denote $S$ as a set such that $(i,j) \in S$ implies $\mathbf{x}_i, \mathbf{x}_j$ are similar. According to~\cite{zhang2018semantic}, the formal formulation of the triplet ranking loss is 
\begin{equation}
\min_{F} \mathcal{L}_{ro} = \sum_{(i,j) \in S, (i,k) \notin S}  g(|\mathbf{h}_i - \mathbf{h}_j|, |\mathbf{h}_i - \mathbf{h}_k|)
\label{eq:trf}
\end{equation}
where $g(\cdot, \cdot)$ has the property such that
\begin{equation}
\begin{split}
g(a,b) &\ge 0 \\
0 \le g(a_2, b) - g(a_1, b) \le a_2 - a_1, \quad a_1 &\le a_2 \\
0 \le g(a, b_1) - g(a, b_2) \le b_2 - b_1, \quad b_1 &\le b_2 \\
\end{split}
\label{eq:g}
\end{equation}

By introducing $C$ auxilary vectors $\mathbf{c}_1, ..., \mathbf{c}_C \in \mathbb{R}^r$ corresponding to $C$ semantic labels, we can arrive at the following triangle inequalities
\begin{equation}
\begin{split}
&|\mathbf{h}_i-\mathbf{h}_j| \le |\mathbf{h}_i-\mathbf{c}_{s}| + |\mathbf{h}_j-\mathbf{c}_{s}|, \quad s \in Y_i \cap Y_j \\
&|\mathbf{h}_i-\mathbf{h}_k| \ge |\mathbf{h}_i-\mathbf{c}_{t}| - |\mathbf{h}_k-\mathbf{c}_{t}|, \quad t \in Y_k, Y_i \cap Y_k = \emptyset
\end{split}
\label{eq:tri}
\end{equation}
and then
\begin{equation}
\begin{split}
g(|\mathbf{h}_i - \mathbf{h}_j|, |\mathbf{h}_i - \mathbf{h}_k|) \le g(|\mathbf{h}_i-\mathbf{c}_{s}|, |\mathbf{h}_i-\mathbf{c}_{t}|) + (|\mathbf{h}_j-\mathbf{c}_{s}|+|\mathbf{h}_k-\mathbf{c}_{t}|) \\
(s \in Y_i \cap Y_j, t \in Y_k, Y_i \cap Y_k = \emptyset)
\end{split}
\label{eq:triest}
\end{equation}

Considering the class labels are balanced and uncorrelated, ~\cite{zhang2018semantic} arrives at the following Unary Upper Bound (UUB) for triplet ranking loss:
\begin{equation}
\begin{split}
\mathbb{E}[\mathcal{L}_{ro}] \le M_{ro} \mathcal{L}_{uo} \quad &\mathcal{L}_{uo} = \sum_{i=1}^n [q(|Y_i|) \sum_{s \in Y_i} l_{c}(\mathbf{h}_i, s) + u(|Y_i|) \sum_{s \in Y_i} |\mathbf{h}_i - \mathbf{c}_s|] \\
&l_{c}(\mathbf{h}_i, s) = -\log \frac{\exp (-|\mathbf{h}_i-\mathbf{c}_{s}|)}{\sum_{j=1}^C \exp(-|\mathbf{h}_i-\mathbf{c}_{j}|)} 
\end{split}
\label{eq:uub}
\end{equation}
where $|Y_i|$ denotes the number of labels $\mathbf{x}_i$ contains. More specifically, for multiclass dataset where $|Y_i|=1$, if the class labels are evenly distributed, we have $M_{ro} = (\frac{n}{C})^2(C-1), q(|Y_i|)=1, u(|Y_i|)=2$. For multilabel dataset where the probability $\mathbb{P}(l \in Y_i) = p$ for all $l=1,2,...,C$, we have $M_{ro} = (C-1)p^2 n^2, q(x)=\frac{C-x}{C-1}(1-p)^x, u(x)=q(x)+(1-p)^2(1-p^2)^{C-2}$. In practical applications, the UUB is very loose, thus we can set $u(|Y_i|)$ to a relatively small value for training the hashcodes, and we name $\mathcal{L}_{uo}$ as {\em Semantic Cluster Unary Loss (SCUL)}.

Eq. (\ref{eq:uub}) establishes the relationship between the triplet ranking loss and the (multilabel) softmax loss, which is easy to implement. Training with SCUL achieves state-of-the-art for single-modal hashing problems.

\subsection{Improved Unary Upper Bound for Structured Multilabel Data}
\label{sec:improve_uub}

\begin{algorithm}[t]
\caption{Determine the coefficients of Eq. (\ref{eq:ot_mtuub_mid})}
\label{alg:determine}
\begin{algorithmic}[1]
\STATE \textbf{Input:} The label vectors $\mathbf{Y} \in \{0,1 \}^{n \times C}$. $\quad // \mathbf{Y}[i,s] = 1 \Leftrightarrow s \in Y_i$
\STATE \textbf{Output:} The coefficients $q_{ist}, u_{is}$ of Eq. (\ref{eq:ot_mtuub_mid}).
\STATE Sample $l$ anchor data from the training data $\mathbf{a}$;
\STATE Initialize $\mathbf{Q}_i = \mathbf{0}_{C \times C} (i=1,...,n), \mathbf{U} = \mathbf{0}_{n \times C}$;
\STATE Compute $\mathbf{Y}' = \mathbf{Y} / \mathrm{sum}(\mathbf{Y}, 1)$; $\quad // \mathrm{sum}(\mathbf{Y},1)$ denotes the matrix sum of $\mathbf{Y}$ along the first axis.
\STATE Compute the similarity matrix $\mathbf{S} = \mathbf{(Y} \mathbf{Y}[\mathbf{a}]^\top>0)$; $\quad // \mathbf{a}$ denotes the rows of $\mathbf{Y}$.
\FOR {$i$ = 1 to $n$}
\STATE Get positive and negative data of $\mathbf{x}_i$ from the anchor set, denote $\mathbf{p}_i, \mathbf{n}_i$; $\quad // (i,j) \in S, (i,k) \notin S, j \in \mathbf{p}_i, k \in \mathbf{n}_i, j,k \in \mathbf{a}$
\STATE Compute $\mathbf{L}_i^s = \mathbf{Y}[i] \circ \mathbf{Y}[\mathbf{p}_i]$; $\quad // \circ$ denotes the element product of the matrices.
\STATE Compute $\mathbf{L}_i^{s'} = \mathbf{L}_i^s / \mathrm{sum}(\mathbf{L}_i^s, 1)$;
\STATE Compute $\mathbf{L}_i^t = \mathbf{Y}'[\mathbf{n}_i]$;
\STATE Compute $\mathbf{Q}_i = \mathrm{sum}(\mathbf{L}_i^{s'}, 0)^\top \mathrm{sum}(\mathbf{L}_i^t, 0)$, and then $q_{ist} = \mathbf{Q}_i[s,t]$
\ENDFOR
\FOR {$i$ in $\mathbf{a}$}
\STATE Get positive and negative data of $\mathbf{x}_i$ from the training set, denote $\mathbf{p}_i, \mathbf{n}_i$; $\quad // \mathbf{p}_i, \mathbf{n}_i \subseteq \{1,...,n\}$
\STATE Compute $\mathbf{L}_i^s = \mathbf{Y}[i] \circ \mathbf{Y}[\mathbf{p}]$;
\STATE Compute $\mathbf{L}_i^{s'} = \mathbf{L}_i^s / \mathrm{sum}(\mathbf{L}_i^s, 1)$;
\STATE $U[\mathbf{p}_i] \gets U[\mathbf{p}_i] + |\mathbf{n}_i| \mathbf{L}_i^{s'}$
\STATE $U[\mathbf{n}_i] \gets U[\mathbf{n}_i] + |\mathbf{p}_i| \mathbf{Y}'[\mathbf{n}_i]$
\ENDFOR
\RETURN $q_{ist} = \mathbf{Q}_i[s,t] \cdot (\frac{n}{l})^2, u_{is} = \mathbf{U}[i,s] \cdot \frac{n}{l}$
\end{algorithmic}
\end{algorithm}

It is clear that the above SCUL in Eq. (\ref{eq:uub}) holds under very harsh conditions such that the labels must be balanced and uncorrelated. However, for structured multilabel dataset, some labels are correlated, in which a data point with a certain label is likely to have other certain labels, making the SCUL inaccurate. Thus the should consider the SCUL for the structured multilabel data.

It is easy to arrive at the following inequality according to Eq. (\ref{eq:triest}):
\begin{equation}
\begin{split}
&g(|\mathbf{h}_i - \mathbf{h}_j|, |\mathbf{h}_i - \mathbf{h}_k|) \\
\le & \sum_{t \in Y_k} \frac{1}{|Y_k|} [g(|\mathbf{h}_i-\mathbf{c}_{s}|, |\mathbf{h}_i-\mathbf{c}_{t}|) + (|\mathbf{h}_j-\mathbf{c}_{s}|+|\mathbf{h}_k-\mathbf{c}_{t}|) ] \\
&\qquad \qquad \qquad \qquad (s \in Y_i \cap Y_j, t \in Y_k, Y_i \cap Y_k = \emptyset)
\end{split}
\label{eq:ml_triest}
\end{equation}

Eq. (\ref{eq:ml_triest}) implies that by the simple permutation and combination, the UUB in Eq. (\ref{eq:trf}) only involves the term $g(|\mathbf{h}_i-\mathbf{c}_{s}|, |\mathbf{h}_i-\mathbf{c}_{t}|), s \in Y_i, t \notin Y_i$ and $|\mathbf{h}_i-\mathbf{c}_{s}|, s \in Y_i$ for all $i=1,2,...,n$. Thus we can arrive the following form of UUB for structured multilabel data such that
\begin{equation}
\mathcal{L}_{ro} \le \sum_{i=1}^n [\sum_{s \in Y_i} \sum_{t \notin Y_i} q_{ist} g(|\mathbf{h}_i - \mathbf{c}_s|, |\mathbf{h}_i - \mathbf{c}_t|) + \sum_{s\in Y_i} u_{is} |\mathbf{h}_i - \mathbf{c}_s|]
\label{eq:ot_mtuub_mid}
\end{equation}

Directly determining the coefficients $q_{ist}, u_{is}$ is almost intractable as it involves $O(n^3)$ triplets for computation. Inspired by the derivation in~\cite{zhang2018semantic}, we propose a novel $O(n)$ strategy to estimate the coefficients by random sampling $l \ll n$ anchor data from the training set, which is shown in Algorithm \ref{alg:determine}. It should be noticed that the coefficients can be accurately computed if regarding all training data as the anchor data. The brief derivations for the algorithm is shown in the supplementary material.

If $g(|\mathbf{h}_i - \mathbf{c}_s|, |\mathbf{h}_i - \mathbf{c}_t|) = l_c(\mathbf{h}_i, s)$ for $ i \in \{1,...,n\}, t \in \{1,...,C \}$ where $l_c(\mathbf{h}_i, s)$ is defined the same as Eq. (\ref{eq:uub}), we can arrive at the improved UUB such that
\begin{equation}
\mathcal{L}_{ro} \le \mathcal{L}_{uo}' = \sum_{i=1}^n [\sum_{s \in Y_i} q_{is} l_c(\mathbf{h}_i, s) + \lambda \sum_{s\in Y_i} u_{is} |\mathbf{h}_i - \mathbf{c}_s|]
\label{eq:ot_mtuub}
\end{equation}
where $q_{is} = \sum_{t \notin Y_i} q_{ist}$ and $\lambda = 1$. As discussed in ~\cite{zhang2018semantic}, the upper bound in Eq. (\ref{eq:ot_mtuub_mid}) is relatively loose, thus we can set $\lambda$ in Eq. (\ref{eq:ot_mtuub}) to a relatively small value for ease of training the hashcodes.

Eq. (\ref{eq:ot_mtuub}) and Algorithm \ref{alg:determine} propose a new loss for training with structured multilabel data where the labels are usually unbalanced and correlated. It should be noticed that we can also apply Eq. (\ref{eq:ot_mtuub}) and Algorithm \ref{alg:determine} to multiclass dataset, in which we just need to sample one data point per class to construct the anchor set. However, directly using Eq. (\ref{eq:uub}) is preferred as the labels can be balanced by sampling the data points.

\subsection{Unary Upper Bound for Cross-modal Learning}
\label{sec:uub_cross}

In the previous subsection, we mainly discuss hashing on single-modal dataset. However, in practical applications, there exists data with cross modalities (images and texts, etc.) and the cross-modal retrieval problems should be considered. Cross-modal hashing is widely applied in which the hashcodes of each modality should be computed for cross-modal retrieval.

The typical triplet loss for cross-modal learning~\cite{zhang2018attention} is denoted as
\begin{equation}
\begin{split}
    \mathcal{L}_{r12} &= \sum_{(i,j) \in S, (i,k) \notin S} g(|\mathbf{h}_{i1} - \mathbf{h}_{j2}|, |\mathbf{h}_{i1} - \mathbf{h}_{k2}|) \\
    \mathcal{L}_{r21} &= \sum_{(i,j) \in S, (i,k) \notin S} g(|\mathbf{h}_{i2} - \mathbf{h}_{j1}|, |\mathbf{h}_{i2} - \mathbf{h}_{k1}|) \\
    \mathcal{L}_{r} &= \mathcal{L}_{r12} + \mathcal{L}_{r21}
\end{split}
\end{equation}
where $g(\cdot, \cdot)$ is defined in Eq. (\ref{eq:g}), $\mathcal{L}_{r12}$ is the triplet loss in which the 1st modality is the query and the 2nd modality is the database, and $\mathcal{L}_{r21}$ is inversed. The distances of codes between cross modalities are involved in the triplet loss as we need to compute the corss-modal distances for retrieval.

By introducing $C$ auxilary vectors $\mathbf{c}_1, ..., \mathbf{c}_C \in \mathbb{R}^r$, we can use the traingle inequality to perform the following distance estimation:
\begin{equation}
\begin{split}
&|\mathbf{h}_{i1}-\mathbf{h}_{j2}| \le |\mathbf{h}_{i1}-\mathbf{c}_{s}| + |\mathbf{h}_{j2}-\mathbf{c}_{s}|, \quad s \in Y_i \cap Y_j \\
&|\mathbf{h}_{i1}-\mathbf{h}_{k2}| \ge |\mathbf{h}_{i1}-\mathbf{c}_{t}| - |\mathbf{h}_{k2}-\mathbf{c}_{t}|, \quad s \in Y_i, t \in Y_k, Y_i \cap Y_k = \emptyset
\end{split}
\end{equation}
and then we can arrive at the following upper bound for the cross-modal triplet loss according to Eq. (\ref{eq:g}):
\begin{equation}
\begin{split}
g(|\mathbf{h}_{i1} - \mathbf{h}_{j2}|, |\mathbf{h}_{i1} - \mathbf{h}_{k2}|) \le g(|\mathbf{h}_{i1}-\mathbf{c}_{s}|, |\mathbf{h}_{i1}-\mathbf{c}_{t}|) \\
+ (|\mathbf{h}_{j2}-\mathbf{c}_{s}|+|\mathbf{h}_{k2}-\mathbf{c}_{t}|) \\
g(|\mathbf{h}_{i2} - \mathbf{h}_{j1}|, |\mathbf{h}_{i2} - \mathbf{h}_{k1}|) \le g(|\mathbf{h}_{i2}-\mathbf{c}_{s}|, |\mathbf{h}_{i2}-\mathbf{c}_{t}|) \\
+ (|\mathbf{h}_{j1}-\mathbf{c}_{s}|+|\mathbf{h}_{k1}-\mathbf{c}_{t}|) \\
(s \in Y_i \cap Y_j, t \in Y_k, Y_i \cap Y_k = \emptyset)
\end{split}
\label{eq:cm_triest}
\end{equation}

The derivations of Eq. (\ref{eq:cm_triest}) are the same as Eq. (\ref{eq:triest}), which can be found in~\cite{zhang2018semantic}. We then have the UUB for cross-modal triplet loss such that
\begin{equation}
\begin{split}
\mathcal{L}_{r12} &\le \mathcal{L}_{u12} = \sum_{i=1}^n [\sum_{s \in Y_i} q_{is} l_c(\mathbf{h}_{i1}, s) + \lambda \sum_{s\in Y_i} u_{is} |\mathbf{h}_{i2} - \mathbf{c}_s|] \\
\mathcal{L}_{r21} &\le \mathcal{L}_{u21} = \sum_{i=1}^n [\sum_{s \in Y_i} q_{is} l_c(\mathbf{h}_{i2}, s) + \lambda \sum_{s\in Y_i} u_{is} |\mathbf{h}_{i1} - \mathbf{c}_s|] \\
\mathcal{L}_{r} &\le \mathcal{L}_{u} = \mathcal{L}_{u12} + \mathcal{L}_{u21}
\end{split}
\label{eq:cm_uub}
\end{equation}

Where $\lambda$ is relatively small as discussed in Sec. \ref{sec:improve_uub}. We name $\mathcal{L}_u$ in Eq. (\ref{eq:cm_uub}) as Cross-Modal Unary Loss (CMUL). It is clear that CMUL reduces the complexity of cross-modal triplet loss to $O(n)$ by introducing $C$ semantic cluster centers $\mathbf{c}_1, ..., \mathbf{c}_C$. Training with $\mathcal{L}_u$ is easy to be implemented, as we can train the two modalities separately with $\mathcal{L}_u$. For the first modality, we can mimimize the distance between $\mathbf{h}_i$ and the corresponding cluster centers $\mathbf{c}_s, s\in Y_i$ with $l_c(\mathbf{h}_{i1}, s)$ and $|\mathbf{h}_{i1} - \mathbf{c}_s|$, and maximize the distances between $\mathbf{h}_{i1}$ and irrelavant centers $\mathbf{c}_t, t \notin Y_i$ by minimizing $l_c(\mathbf{h}_{i1}, s)$. The same procedure holds in the second modality. Meanwhile, the hashcodes of two modalities are expected to be aligned by clustering around the same semantic centers $\mathbf{c}_1, ..., \mathbf{c}_C$.

\begin{figure}[t]
    \setlength{\abovecaptionskip}{2pt}
    \setlength{\belowcaptionskip}{0pt}
    \centering
    \includegraphics[scale=0.64]{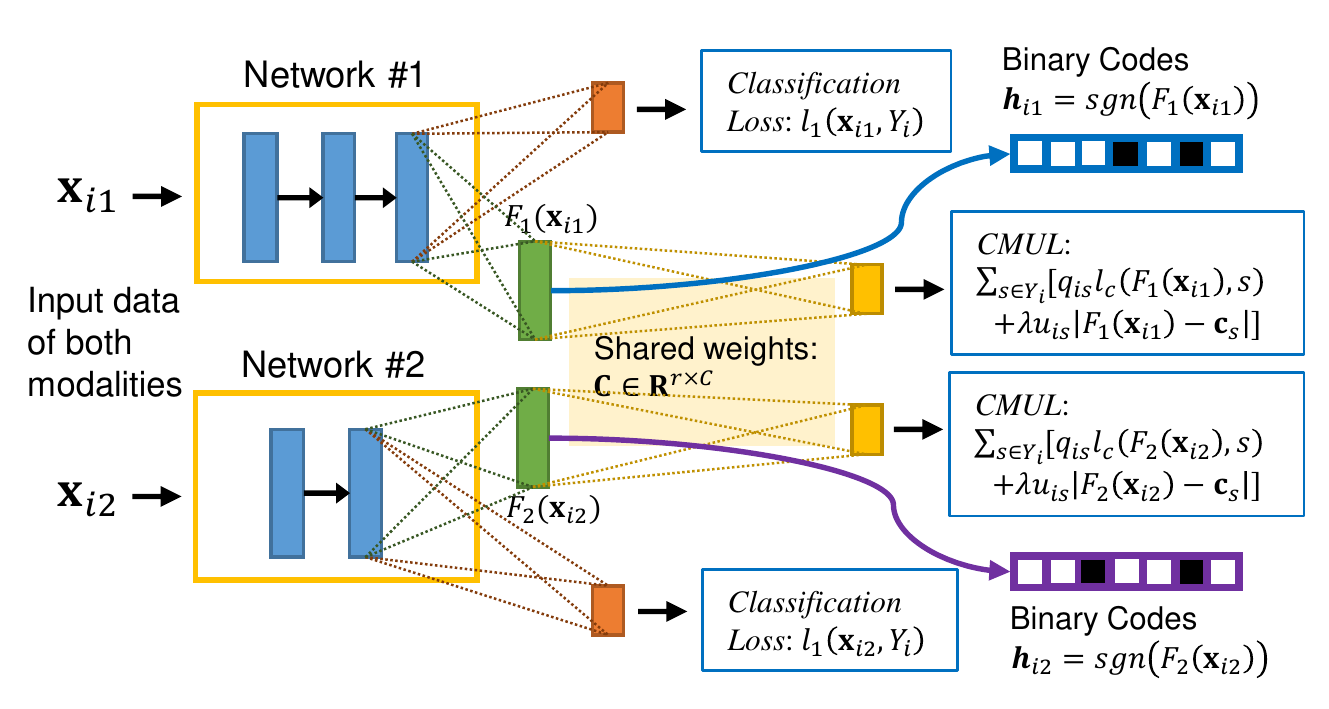}
    \caption{Illustration on the Joint Cluster Cross-modal Hashing (JCCH) algorithm. For each modality, the hash codes can be obtained from the hashing layer (in green rectangle). The network is jointly trained by the CMUL, the quantization loss and the multi-label softmax loss, defined in Eq. (\ref{eq:mc_all}). The network of each modality can be trained almost separately, except that the parameters of the last fully-connected layer before the CMUL, denote $\mathbf{C} \in \mathbb{R}^{r \times C}$ (in light yellow), are shared in both modalities.}
    \label{fig:jcch_framework}
\end{figure}

\section{Joint Cluster Cross-modal Hashing}

In this section, we propose a novel and efficient cross-modal deep hashing algorithm, called Joint Cluster Cross-modal Hashing (JCCH), in which the CMUL is applied. As the complexity of CMUL is $O(n)$, the proposed algorithm is expected to be efficient.

\subsection{Overall Architecture}

The overall architecture is shown in Figure \ref{fig:jcch_framework}, which is similar with~\cite{zhang2018semantic} if just considering the individual modality. For each modality, the base network is the classification network (e.g. AlexNet for the image modality and language model for the text modality). After the last but one layer in the classification network of each modality, we concatenate two fully-connected layers without non-linear activation. The first layer denotes the hashing layer and the number of activations is $r$. The second layer has $C$ outputs with parameter $\mathbf{C} = [\mathbf{c}_1, ..., \mathbf{c}_C] \in \mathbb{R}^{r \times C}$, which is the semantic cluster centers defined in Eq. (\ref{eq:cm_uub}). Note that the networks of both modalities share the same $\mathbf{C}$ to minimize the gap between the two modalities.

Denote $F_1(\mathbf{x}_{i1}), F_2(\mathbf{x}_{i2}) \in \mathbb{R}^r, i=1,2,...,n$ as the activations of the hashing layer in each modality, the hashcodes of both modalities can be obtained with $\mathbf{h}_{i1} = \mathrm{sgn}(F_1(\mathbf{x}_{i1})), \mathbf{h}_{i2} = \mathrm{sgn}(F_2(\mathbf{x}_{i2}))$, where $\mathrm{sgn}$ is the element-wise sign function. To obtain a good hash function, we should optimize $F_1, F_2, \mathbf{C}$ with the loss based on CMUL.

\subsection{Loss Function and Relaxation}

We should optimize the CMUL such that
\begin{equation}
\begin{split}
\min_{F_1, F_2, \mathbf{C}} \mathcal{L}_{u} &= \mathcal{L}_{u12} + \mathcal{L}_{u21} \\
&= \sum_{i=1}^n \Big\{ \sum_{s \in Y_i} \big[ q_{is} l_c(\mathbf{h}_{i1}, s)
+ \lambda u_{is} |\mathbf{h}_{i1} - \mathbf{c}_s| \\
& \qquad \qquad + q_{is} l_c(\mathbf{h}_{i2}, s) + \lambda u_{is} |\mathbf{h}_{i2} - \mathbf{c}_s|\big] \Big \}
\end{split}
\label{eq:cm_hash_loss}
\end{equation}
where $q_{is}, u_{is}$ are computed with Algorithm \ref{alg:determine} for the multilabel case. The computed $q_{is}, u_{is}$ are large because there are large number of data triplets. For ease of training, we rescale $q_{is}, u_{is}$ such that the mean of nonzero value of $q_{is}$ is 1:
\begin{equation}
M = \frac{\sum_{i,s}{q_{is}}}{\sum_{i}{|Y_i|}}, \quad q_{is} \gets q_{is}/M, \quad u_{is} \gets u_{is} / M 
\end{equation}

Similar as ~\cite{zhang2018semantic}, we add another classification loss for each modality such that $\mathcal{L}_{1} = \mathcal{L}_{11} + \mathcal{L}_{12} = \sum_{i=1}^n l_{1}(\mathbf{x}_{i1}, \mathbf{Y}_i) + \sum_{i=1}^n l_{1}(\mathbf{x}_{i2}, \mathbf{Y}_i)$ for faster training convergence, where $l_1(\cdot, \cdot)$ is the classification (multilabel) softmax loss to optimize the base classification network for each modality. Then the overall loss to be optimized is 
\begin{equation}
\min_{F_1, F_2, \mathbf{C}} \mathcal{L} = \mathcal{L}_u + \mu \mathcal{L}_1
\label{eq:mc_l}
\end{equation}

It is clear that directly optimizing Eq. (\ref{eq:mc_l}) is intractable. Followed by the previous works~\cite{li2015feature,cao2016deep}, we remove the $\mathrm{sgn}$ function and add the quantization loss. As the norms of two modalities may diverse and the norm of real-valued vector before the $\mathrm{sgn}$ function does not affect the generated binary codes, we do not necessarily constrain the norm of real-valued features but just push the elements away from zero. Therefore, we use the quantization loss proposed in~\cite{zhang2018semantic} such that $l_q(\mathbf{f}) = 1 - \frac{\mathbf{1}^\top \mathbf{f}}{\| \mathbf{1} \|_{1.5} \| \mathbf{f} \|_{3}}$ where $\| \cdot \|_p$ is the $p$-norm. We then have the following relaxed problem:
\begin{equation}
\begin{split}
\min_{F_1, F_2, \mathbf{C}} \mathcal{L}^r &= \mathcal{L}_u^r + \mu \mathcal{L}_1 + \alpha \sum_{i=1}^n \big[ l_q(F_1(\mathbf{x}_{i1})) + l_q(F_2(\mathbf{x}_{i2})) \big] \\
\mathcal{L}_u^r &= \sum_{i=1}^n \Big\{ \sum_{s \in Y_i} \big[ q_{is} l_c(F_1(\mathbf{x}_{i1}), s)
+ \lambda u_{is} |F_1(\mathbf{x}_{i1}) - \mathbf{c}_s| \\
& \qquad \qquad + q_{is} l_c(F_2(\mathbf{x}_{i2}), s) + \lambda u_{is} |F_2(\mathbf{x}_{i2}) - \mathbf{c}_s|\big] \Big \}
\end{split}
\label{eq:mc_relax}
\end{equation}

Moreover, if the two modalities are paired, (i.e., both modalities correspond to the same data point), we can force the corresponding hashcodes to be similar to make the codes of two modalities better aligned. Then the final loss to be optimized is
\begin{equation}
\begin{split}
\min_{F_1, F_2, C} \mathcal{L}^r = \mathcal{L}_u^r + \mu \mathcal{L}_1 +  \alpha \sum_{i=1}^n \big[ l_q(F_1(\mathbf{x}_{i1})) + l_q(F_2(\mathbf{x}_{i2})) \big] \\
+ \beta \sum_{i=1}^n (1-\cos<F_1(\mathbf{x}_{i1}), F_2(\mathbf{x}_{i2})>)
\end{split}
\label{eq:mc_all}
\end{equation}

It should be noticed that if two modalities are not paired, we can simply set $\beta = 0$.

\subsection{Training Details}

From Eq. (\ref{eq:mc_all}) we can see that the proposed JCCH is easy to implement, which can be trained end-to-end by gradient descent with back-propagation. To prevent the training procedure collapsed in which $F_1(\cdot), F_2(\cdot)$ and the semantic cluster centers $\mathbf{c}_i$ go to zero, we follow~\cite{zhang2018semantic} such that the centers are initialized so that the norms are relatively large. 

\section{Experiments}

In this section, we conduct various types of large-scale cross-modal retrieval experiments to show the efficiency of the proposed JCCH method. We study the performance and training efficiency of the JCCH compared with recent state-of-the-art cross-modal hashing methods. The ablation study on the efficiency of improved UUB is also conducted in this section.

\subsection{Datasets and Evaluation Metrics}

We run large-scale retrieval experiments on various types of cross-modal datasets. First, we conduct experiments on two sketch-image datasets: TU-Berlin~\cite{eitz2012humans} Extension and Sketchy~\cite{sangkloy2016sketchy}. TU-Berlin Extension contains 20,000 free-hand sketches and 204,489 natural images from 250 categories. Sketchy consists of 75,471 hand-drawn images and 73,002 natural images from 125 categories~\cite{liu2017deep}. Second, we use three text-image datasets for evaluation: MIRFLIKR-25K, NUS-WIDE and IAPR TC-12. MIRFLIKER-25K contains 25,000 image-text pairs and each point is annotated with 24 labels. NUS-WIDE consists of about 270K images and each image associates with 81 ground truth concept labels. Following~\cite{jiang2017deep}, we only use the images associated with 21 most frequent concepts, containing about 190K images. IAPR TC-12 dataset consists of 20,000 image-text pairs with 255 labels. For text data, we simply follow~\cite{jiang2017deep} and use bag-of-words vector, where the dimension is 1386 in MIRFLIKER-25K, 1000 in NUS-WIDE and 3529 in IAPR TC-12.

The experimental protocols are the same as that in~\cite{jiang2017deep,liu2017deep}. For sketch-image retrieval, we randomly select 2,500 sketches (10 per class) in TU-Berlin and 6250 sketches (50 per class) in Sketchy as the query set, and the rest images and sketches are regarded as the training set and the retrieval database. For text-image retrieval, we randomly select 2,000 data points in MIR-FLICKER and IAPR TC-12 dataset and 2,100 points in NUS-WIDE as the query set. The rest data form the retrieval database. Moreover, we randomly take 10,000 points in MIR-FLIKER and IAPR TC-12 dataset and 10,500 points in NUS-WIDE from the retrieval database to build the training set. The data pairs are similar if they share one or more semantic labels.

Our JCCH algorithm is implemented with PyTorch\footnote{http://pytorch.org} framework. For image and sketch modality, we use the pre-trained deep networks (e.g. AlexNet, VGGNet, ResNet) for initialization, and the images/sketches are resized to $224 \times 224$ to feed the deep network. For text modality, we follow~\cite{jiang2017deep} in which a two-layer MLP with ReLU activation is applied for training text vectors. The dimension of hidden layer is 8192. We follow~\cite{zhang2018semantic} in that the last classification layer and the hashing layer are initialized by "Gaussian" initializer with zero mean and standard deviation 0.01 for both modalities, and the semantic cluster centers $\mathbf{C}$ are initialized with standard deviation 0.5. 

For training the network, we use SGD for optimization with momentum 0.9 and weight decay 0.0001. For image modality, the initial learning rate is set to 0.001 before the last but one layer and 0.01 for the rest layer. For text or sketch modality, the learning rate is set to 0.01-0.03 for all layers. The hyper-parameters $\lambda, \mu, \alpha, \beta$ is different according to datasets, which are determined according to the validation set. We choose $\lambda=0.001$ in two sketch datasets, and set $\{ \lambda=0.04, \beta=0 \}$ in MIR-FLIKER, $\{ \lambda=0.002, \beta=0.2 \}$ in NUS-WIDE and $\{ \lambda=0.0002, \beta=0.1 \}$ in IAPR TC-12. Inspired by~\cite{zhang2018semantic}, $\mu$ is set to 0.1 and $\alpha$ is set such that the quantization loss of both modalities is around 0.15. As the training set is not too large, we regard all the training data as the anchor data to compute the coefficients $q_{is}, u_{is}$ in Algorithm \ref{alg:determine}. Sec. \ref{sec:ablation} will discuss the size of anchor data $l$ in detail. The training is down on a server with two Intel(R) Xeon(R) E5-2683 v3@2.0GHz CPUs, 256GB RAM and Geforce GTX TITAN Pascal with 12GB memory. 

The evaluation protocols are the same as~\cite{jiang2017deep,liu2017deep}. For sketch-image retrieval, we report the mean average precision (MAP) and precision at top-200 retrieval candidates. For image-text retrieval, we report the compared results in terms MAP, etc. Note that we report the MAP value on all retrieved data points. We also perform ablation study to show the effectiness of improved UUB on the structured multilabel data. Each experiment is run for 5 times and get the average result. 

\begin{table*}[]
    \setlength{\abovecaptionskip}{2pt}
    \setlength{\belowcaptionskip}{0pt}
    \centering
    \small
    \begin{tabular}{|c|c|ccc|ccc||ccc|ccc|}
        \hline
        \multirow{3}{*}{Method} & \multirow{3}{*}{Feature} & \multicolumn{6}{c||}{TU-Berlin Extension} & \multicolumn{6}{c|}{Sketchy} \\
        \cline{3-14}
         & & \multicolumn{3}{c|}{MAP} & \multicolumn{3}{c||}{Precision@200} & \multicolumn{3}{c|}{MAP} & \multicolumn{3}{c|}{Precision@200} \\
         \cline{3-14}
         & & 32 bits & 64 bits & 128 bits & 32 bits & 64 bits & 128 bits & 32 bits & 64 bits & 128 bits & 32 bits & 64 bits & 128 bits \\
         \hline
         CVH~\cite{liu2017deep} & \multirow{3}{*}{hand-craft} & 0.214 & 0.294 & 0.318 & 0.305 & 0.411 & 0.449 & 0.325 & 0.525 & 0.624 & 0.459 & 0.641 & 0.773 \\
         SePH~\cite{liu2017deep} & & 0.198 & 0.270 & 0.282 & 0.307 & 0.380 & 0.398 & 0.534 & 0.607 & 0.640 & 0.694 & 0.741 & 0.768 \\
         CCA~\cite{liu2017deep} & & 0.276 & 0.366 & 0.365 & 0.333 & 0.482 & 0.536 & 0.361 & 0.555 & 0.705 & 0.379 & 0.610 & 0.775 \\
         \hline
         DCMH~\cite{liu2017deep} & \multirow{3}{*}{AlexNet} & 0.274 & 0.382 & 0.425 & 0.332 & 0.467 & 0.540 & 0.560 & 0.622 & 0.656 & 0.730 & 0.771 & 0.784 \\
         DSH~\cite{liu2017deep} & & 0.358 & 0.521 & 0.570 & 0.486 & 0.655 & 0.694 & 0.653 & 0.711 & 0.783 & 0.797 & 0.858 & 0.866 \\
         \textbf{JCCH(Ours)} & & \textbf{0.694} & \textbf{0.721} & \textbf{0.739} & \textbf{0.707} & \textbf{0.718} & \textbf{0.726} & \textbf{0.919} & \textbf{0.930} & \textbf{0.931} & \textbf{0.909} & \textbf{0.915} & \textbf{0.920} \\
         \hline
         GDH~\cite{zhang2018generative} & \multirow{2}{*}{ResNet18} & 0.563 & 0.690 & 0.651 & \multicolumn{3}{c||}{N/A} & 0.724 & 0.811 & 0.784 & \multicolumn{3}{c|}{N/A} \\ 
         \textbf{JCCH(Ours)} & & \textbf{0.752} & \textbf{0.771} & \textbf{0.760} & \textbf{0.729} & \textbf{0.739} & \textbf{0.724} & \textbf{0.918} & \textbf{0.927} & \textbf{0.934} & \textbf{0.913} & \textbf{0.918} & \textbf{0.925} \\
         \hline
         \textbf{JCCH(Ours)} & SE-ResNeXt-101 & \textbf{0.811} & \textbf{0.821} & \textbf{0.836} & \textbf{0.775} & \textbf{0.779} & \textbf{0.787} & \textbf{0.950} & \textbf{0.950} & \textbf{0.955} & \textbf{0.944} & \textbf{0.944} & \textbf{0.946} \\
         \hline
    \end{tabular}
    \caption{Compared results of different cross-modal methods on category-level sketch-to-image retrieval problems. The results with citations are directly copied from the corresponding papers. The best results are highlighted in boldface.}
    \label{tab:res_sketch}
\end{table*}

\begin{table*}[]
    \setlength{\abovecaptionskip}{2pt}
    \setlength{\belowcaptionskip}{0pt}
    \centering
    \small
    \begin{tabular}{|c|c|c|ccc|ccc|ccc|}
        \hline
        \multirow{2}{*}{Task} & \multirow{2}{*}{Network} & \multirow{2}{*}{Method} & \multicolumn{3}{c|}{MIRFLIKER-25K} & \multicolumn{3}{c|}{NUS-WIDE} & \multicolumn{3}{c|}{IAPR TC-12} \\
        & & & 16 bits & 32 bits & 64 bits & 16 bits & 32 bits & 64 bits & 16 bits & 32 bits & 64 bits  \\
        \hline
        \multirow{8}{*}{$I \to T$} & \multirow{4}{*}{AlexNet} & DCMH~\cite{jiang2017deep} & 0.741 & 0.747 & 0.749 & 0.590 & 0.603 & 0.609 & 0.453 & 0.473 & 0.484 \\
         & & SSAH~\cite{li2018self} & 0.782 & 0.790 & 0.800 & 0.642 & 0.636 & 0.639 & \multicolumn{3}{c|}{N/A} \\
         & & \textbf{JCCH(Ours)} & 0.763 & 0.785 & \textbf{0.802} & 0.626 & \textbf{0.651} & \textbf{0.665} & \textbf{0.476} & \textbf{0.515} & \textbf{0.552} \\
         \cline{3-12}
         & & \textbf{JCCH+DCMH(Ours)} & \textbf{0.802} & \textbf{0.818} & \textbf{0.825}  & \textbf{0.660} & \textbf{0.670} & \textbf{0.674} & \textbf{0.555} & \textbf{0.579} & \textbf{0.600}  \\
        \cline{2-12}
         & \multirow{4}{*}{VGGNet} & PRDH~\cite{zhang2018attention} & 0.750 & 0.755 & 0.761 & 0.611 & 0.630 & 0.628 & 0.500 & 0.494 & 0.513 \\
         & & SSAH~\cite{li2018self} & \textbf{0.797} & 0.809 & 0.810 & 0.636 & 0.636 & 0.637 & \multicolumn{3}{c|}{N/A} \\
         & & ADAH~\cite{zhang2018attention} & 0.756 & 0.772 & 0.772 & \textbf{0.640} & 0.629 & 0.652 & \textbf{0.529} & \textbf{0.528} & 0.544 \\
         & & \textbf{JCCH(Ours)} & 0.786 & \textbf{0.812} & \textbf{0.820} & \textbf{0.640} & \textbf{0.672} & \textbf{0.687} & 0.485 & 0.526 & \textbf{0.569} \\
        \hline
        \multirow{8}{*}{$T \to I$} & \multirow{4}{*}{AlexNet} & DCMH~\cite{jiang2017deep} & 0.782 & 0.790 & 0.793 & 0.639 & 0.651 & 0.657 & 0.519 & 0.538 & 0.547 \\
         & & SSAH~\cite{li2018self} & 0.791 & 0.795 & 0.803 & 0.669 & 0.662 & 0.666 & \multicolumn{3}{c|}{N/A} \\
         & & \textbf{JCCH(Ours)} & 0.759 & 0.778 & 0.792 & 0.630 & 0.654 & \textbf{0.674} & 0.478 & 0.518 & \textbf{0.557} \\
         \cline{3-12}
         & & \textbf{JCCH+DCMH(Ours)} & \textbf{0.802} & \textbf{0.816} & \textbf{0.821} & \textbf{0.660} & \textbf{0.682} & \textbf{0.687} & \textbf{0.531} & \textbf{0.567} & \textbf{0.588} \\
        \cline{2-12}
         & \multirow{4}{*}{VGGNet} & PRDH~\cite{zhang2018attention} & 0.789 & 0.796 & 0.796 & 0.653 & 0.692 & 0.672 & 0.524 & 0.543 & 0.555 \\
         & & SSAH~\cite{li2018self} & 0.782 & 0.797 & 0.799 & 0.653 & 0.676 & 0.683 & \multicolumn{3}{c|}{N/A} \\
         & & ADAH~\cite{zhang2018attention} & \textbf{0.792} & \textbf{0.806} & \textbf{0.807} & \textbf{0.679} & \textbf{0.697} & \textbf{0.704} & \textbf{0.536} & \textbf{0.557} & \textbf{0.565} \\
         & & \textbf{JCCH(Ours)} & 0.754 & 0.791 & \textbf{0.807} & 0.649 & 0.681 & 0.697 & 0.479 & 0.514 & 0.560 \\
        \hline
    \end{tabular}
    \caption{Compared results of different cross-modal hashing methods on text-image retrieval problems. The results with citations are directly copied from the corresponding papers. The best results are highlighted in boldface.}
    \label{tab:res_cross}
\end{table*}

\begin{figure}
    \setlength{\abovecaptionskip}{2pt}
    \setlength{\belowcaptionskip}{0pt}
    \centering
    \includegraphics[scale=0.2]{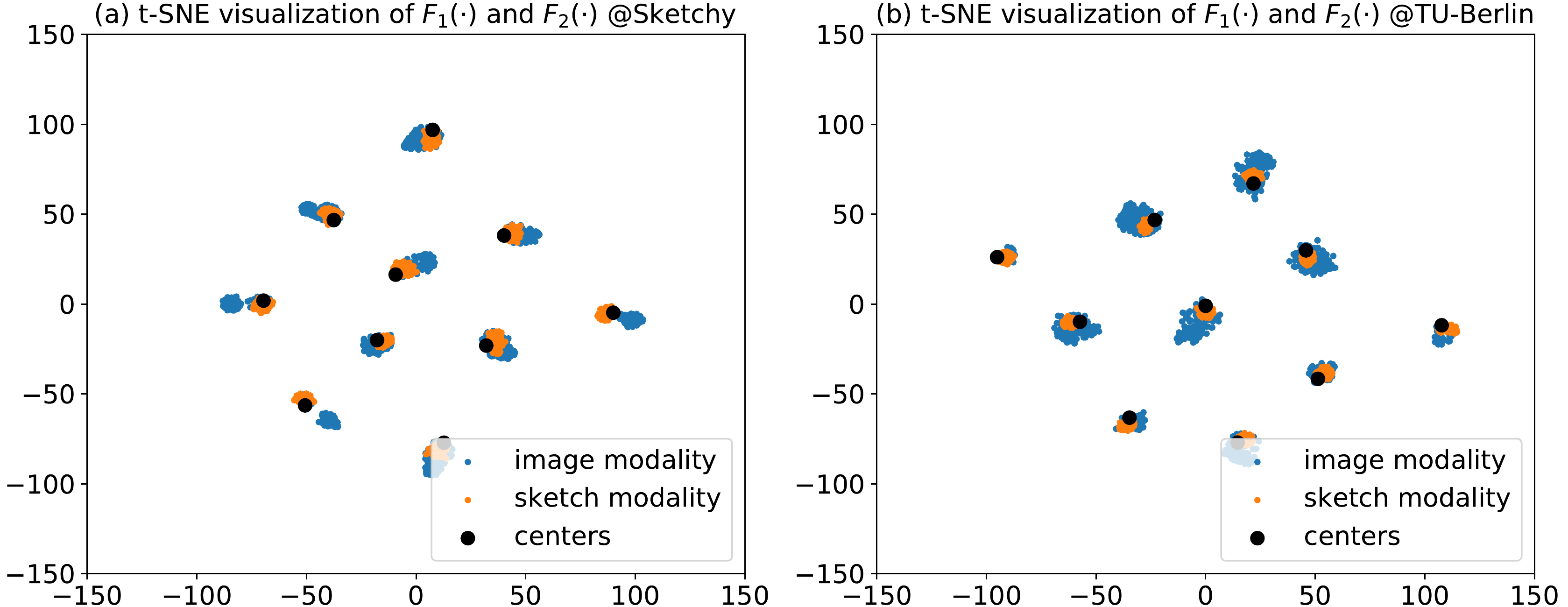}
    \caption{T-SNE visualization on the normalized representations of different modalities $F_1(\cdot), F_2(\cdot)$ . For each modality, we random sample 1,000 images of 10 classes from the training set for visualization, and the corresponding semantic cluster centers are also presented. The code length is 64.}
    \label{fig:vis_sketch}
\end{figure}

\subsection{Results on Category-Level Sketch-Image Retrieval}

We compare the JCCH algorithm with the state-of-the-art deep sketch-to-image retrieval algorithm including DSH~\cite{liu2017deep}, GDH~\cite{zhang2018generative} and other retrieval algorithms like CVH~\cite{kumar2011learning}, SePH~\cite{lin2015semantics}, DCMH~\cite{jiang2017deep}, CCA~\cite{thompson2005canonical}, etc. We do not report other baselines as they performs much inferior than DSH and GDH. As each data point correspond to just one semantic label, we just apply $q_{is} = u_{is} = 1$ in Eq. (\ref{eq:mc_relax}). We conduct experiments with different ImageNet pre-trained models for fair comparison. Results are shown in Table \ref{tab:res_sketch}. With the network structure fixed, the MAP value of our proposed JCCH is greater than the state-of-the-art DSH~\cite{liu2017deep} method by over 29\% in TU-Berlin Extension and over 18\% in Sketchy, and outperforms GDH~\cite{zhang2018generative} by over 11\% in TU-Berlin and 13\% in Sketchy, showing the effectiveness of the proposed JCCH method. It should be noticed that the previous state-of-the-art DSH~\cite{liu2017deep} and GDH~\cite{zhang2018generative} are not end-to-end where the discrete constraints are considered separately, ours is a simple end-to-end algorithm which can be implemented very easily but achieves better performance.

Eq. (\ref{eq:mc_relax}) implies that the learned hashcodes of both modalities can be aligned as $|\mathbf{h}_{i1} - \mathbf{c}_s|, |\mathbf{h}_{i2} - \mathbf{c}_s|$ are expected to be small. To what follows, we also conduct t-SNE visualization on the normalized representations of both modalities $F_{i1}(\cdot), F_{i2}(\cdot)$ from the training data, which is shown in Figure \ref{fig:vis_sketch}. It can be seen clearly that both modalities are aligned by clustering to the shared centers, showing the effectiveness of the proposed CMUL.

\subsection{Results on Text-Image Retrieval and Convergence Analysis}

\begin{figure}[t]
    \setlength{\abovecaptionskip}{2pt}
    \setlength{\belowcaptionskip}{0pt}
    \centering
    \includegraphics[scale=0.18]{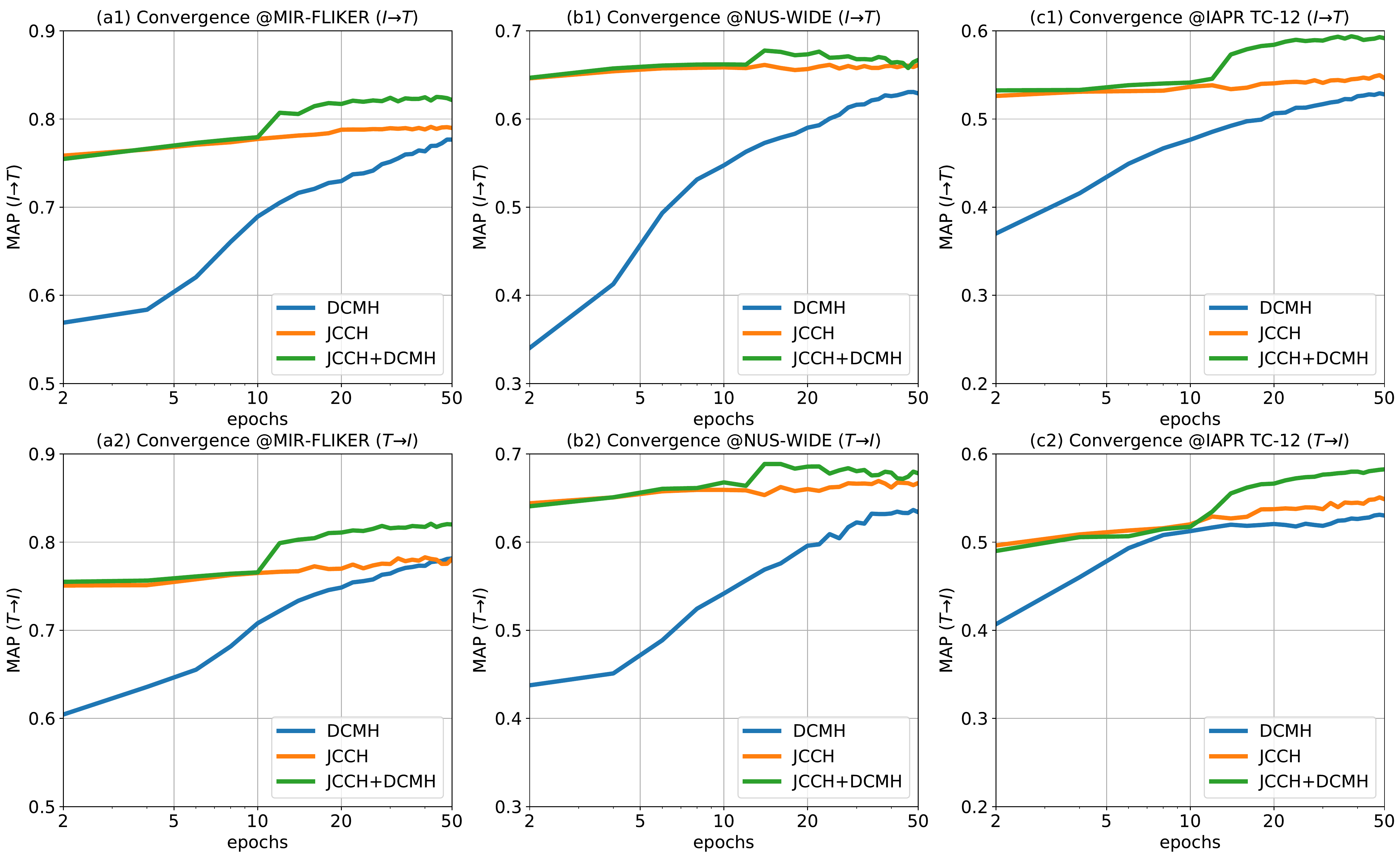}
    \caption{The growth of MAP at different training epochs on text-image retrieval problems. JCCH+DCMH denotes training JCCH for 10 epochs for initialization before training DCMH. The code length is 64.}
    \label{fig:converge}
\end{figure}

We compare our JCCH algorithm with recent state-of-the-art image-text retrieval algorithm including DCMH~\cite{jiang2017deep}, PRDH~\cite{yang2017pairwise}, SSAH~\cite{li2018self}, ADAH~\cite{zhang2018attention}. We do not report other baselines because the settings are different or the results are inferior than the above methods. Note that both image-to-text ($I \to T$) and text-to-image ($T \to I$) retrieval tasks are performed. Results are shown in Table \ref{tab:res_cross}. It can be seen that the proposed JCCH performs better than or comparable with the state-of-the-art baselines on most settings, especially in the image-to-text retrieval ($I \to T$) tasks. It should also be noticed that the proposed JCCH method may perform inferior on some settings. The reason may be that the pairwise or triplet losses directly model the similarity information, while the proposed JCCH methods the similarities indirectly with structured labels.

However, it is clear that the proposed JCCH is expected to train much faster than other baselines as the complexity is just $O(n)$. Figure \ref{fig:converge} shows the MAP value at different training epochs. Note that we re-implement the DCMH~\cite{jiang2017deep} algorithm and the performance is slightly different from the original paper. Compared with DCMH in which the pairwise loss is used, ours is able to converge to the desired results in just a few epochs. It is clear that the proposed JCCH is able to train very fast to achieve desired hashing performance compared with other baselines which adopt pairwise or triplet losses for training.

To what follows, we can adopt the proposed JCCH for initialization before training the state-of-the-art cross-modal hashing algorithms to achieve fast training and best performance simultaneously. Figure \ref{fig:converge} shows the performance of DCMH after training JCCH for 10 epochs, denote JCCH+DCMH. It can also be seen clearly that using JCCH for initialization is able to make the training procedure faster. Moreover, the JCCH initialization procedure is able to achieve much better results by over 0.02 compared with the state-of-the-art baselines. We also conduct extensive image-text retrieval experiments with JCCH+DCMH, and the results are shown in Table \ref{tab:res_cross}. The results shows that JCCH+DCMH achieves state-of-the-art cross-modal hashing performance.

\subsection{The Effectiveness of Improved UUB for Structured Multilabel Data}
\label{sec:ablation}

\begin{table}[]
    \setlength{\abovecaptionskip}{2pt}
    \setlength{\belowcaptionskip}{0pt}
    \centering
    \small
    \begin{tabular}{|c|cc|cc|cc|}
        \hline
        \multirow{2}{*}{Method} & \multicolumn{2}{c|}{MIRFLIKER-25K} & \multicolumn{2}{c|}{NUS-WIDE} & \multicolumn{2}{c|}{IAPR-TC12} \\
         & 32 bits & 64 bits & 32 bits & 64 bits & 32 bits & 64 bits \\
        \hline
        \multicolumn{7}{|c|}{$I \to T$} \\
        \hline
        JCCH-B & 0.780 & 0.785 & 0.642 & 0.647 & 0.487 & 0.535 \\
        JCCH & \textbf{0.785} & \textbf{0.802} & \textbf{0.651} & \textbf{0.665} & \textbf{0.515} & \textbf{0.552} \\
        \hline
        \multicolumn{7}{|c|}{$T \to I$} \\
        \hline
        JCCH-B & 0.770 & 0.773 & 0.647 & 0.660 & 0.495 & 0.540 \\
        JCCH & \textbf{0.778} & \textbf{0.792} & \textbf{0.654} & \textbf{0.674} & \textbf{0.518} & \textbf{0.557} \\
        \hline
    \end{tabular}
    \caption{Compared results on MAP value of variants of JCCH on cross-modal hashing datasets. JCCH-B denotes we directly use $q_{is}=1/|Y_i|, u_{is}=1$ in Eq. (16) (the same loss proposed in~\cite{zhang2018semantic}) for training. The best results are shown in boldface.}
    \label{tab:jcch_variant}
\end{table}

In this section, we study the effectiveness of the improved UUB in Sec. \ref{sec:improve_uub}. We use the SCUL~\cite{zhang2018semantic} as the baseline where we simply set $q_{is} = 1/|Y_i|, u_{is}=1$ for Eq. (\ref{eq:mc_relax}). We name the variant as JCCH-B. Table \ref{tab:jcch_variant} shows the cross-modal image-text retrieval results. It can be seen clearly that adopting the improved UUB is able to improve the retrieval performance of structured multilabel dataset significantly, which implies that the improved UUB can better model the triplet ranking loss for structured multilabel dataset with highly unbalanced and correlated labels.

\begin{figure}[t]
    \setlength{\abovecaptionskip}{2pt}
    \setlength{\belowcaptionskip}{0pt}
    \centering
    \includegraphics[scale=0.18]{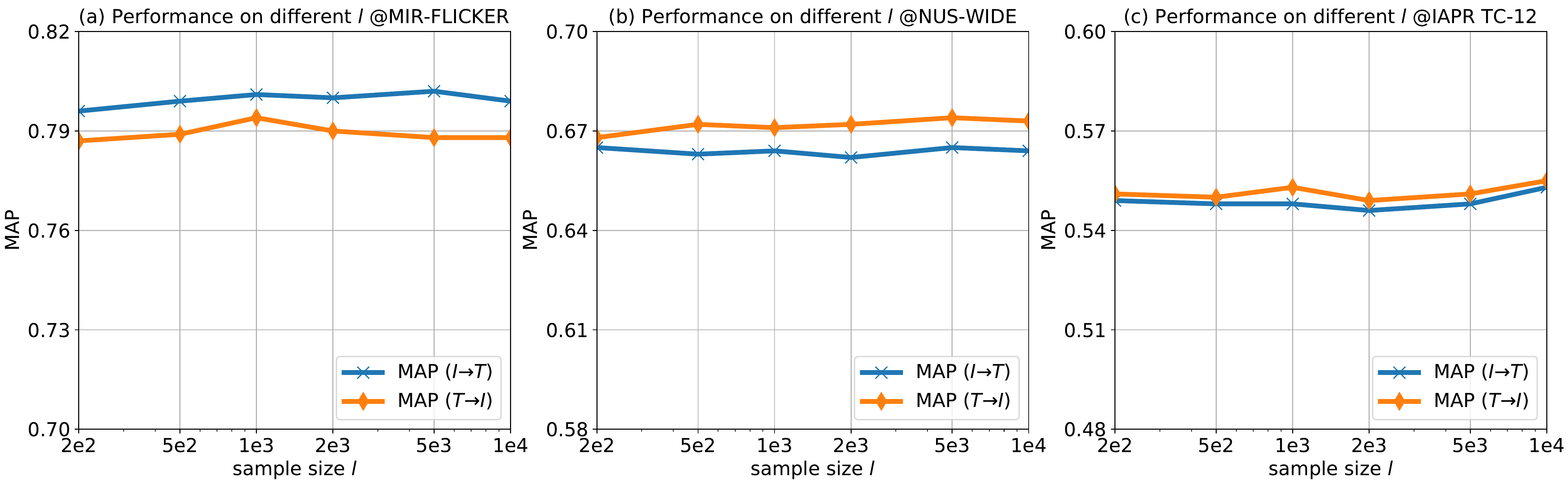}
    \caption{Comparative results of different sample sizes (denote $l$) for determining $q_{is}, u_{is}$ (in Algorithm \ref{alg:determine}) on image-text dataset. The code length is 64.}
    \label{fig:sample_size}
\end{figure}

For large-scale training data, we have to sample a small subset of anchor data to determine the coefficients $q_{is}, u_{is}$. We also conduct experiments of different number of anchor data $l$. Figure \ref{fig:sample_size} shows the results of different sample sizes. It can be seen clearly that small number of anchor data for computing $q_{is}, u_{is}$ have little influence on the hashing performance, and the results are much better than JCCH-B. Thus we are able to choose the number of anchor data freely.

\begin{figure}[t]
    \setlength{\abovecaptionskip}{2pt}
    \setlength{\belowcaptionskip}{0pt}
    \centering
    \includegraphics[scale=0.18]{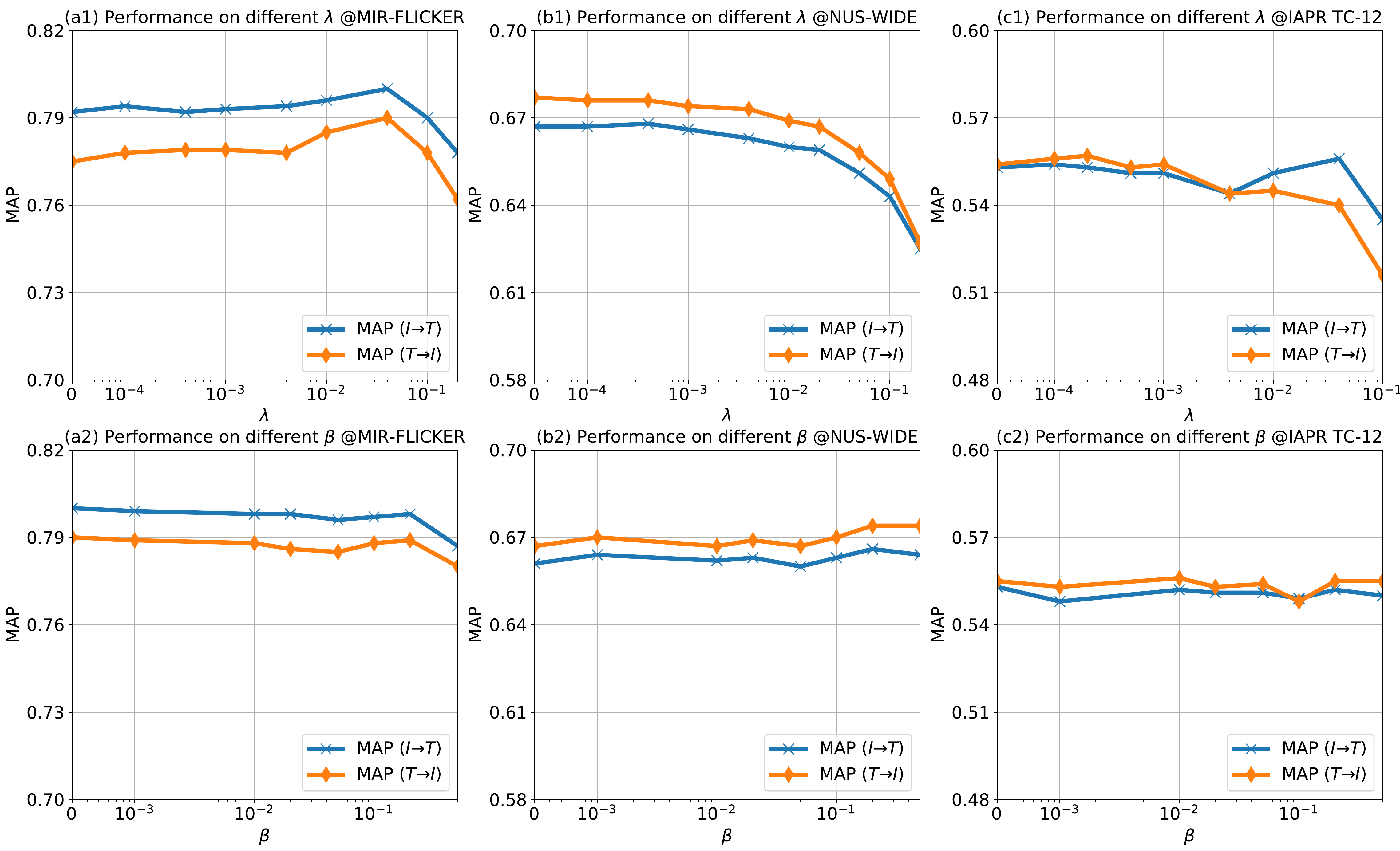}
    \caption{Comparative results in terms of MAP of different hyper-parameters on image-text dataset. The code length is 64.}
    \label{fig:params}
\end{figure}

\begin{figure}[t]
    \setlength{\abovecaptionskip}{2pt}
    \setlength{\belowcaptionskip}{0pt}
    \centering
    \includegraphics[scale=0.18]{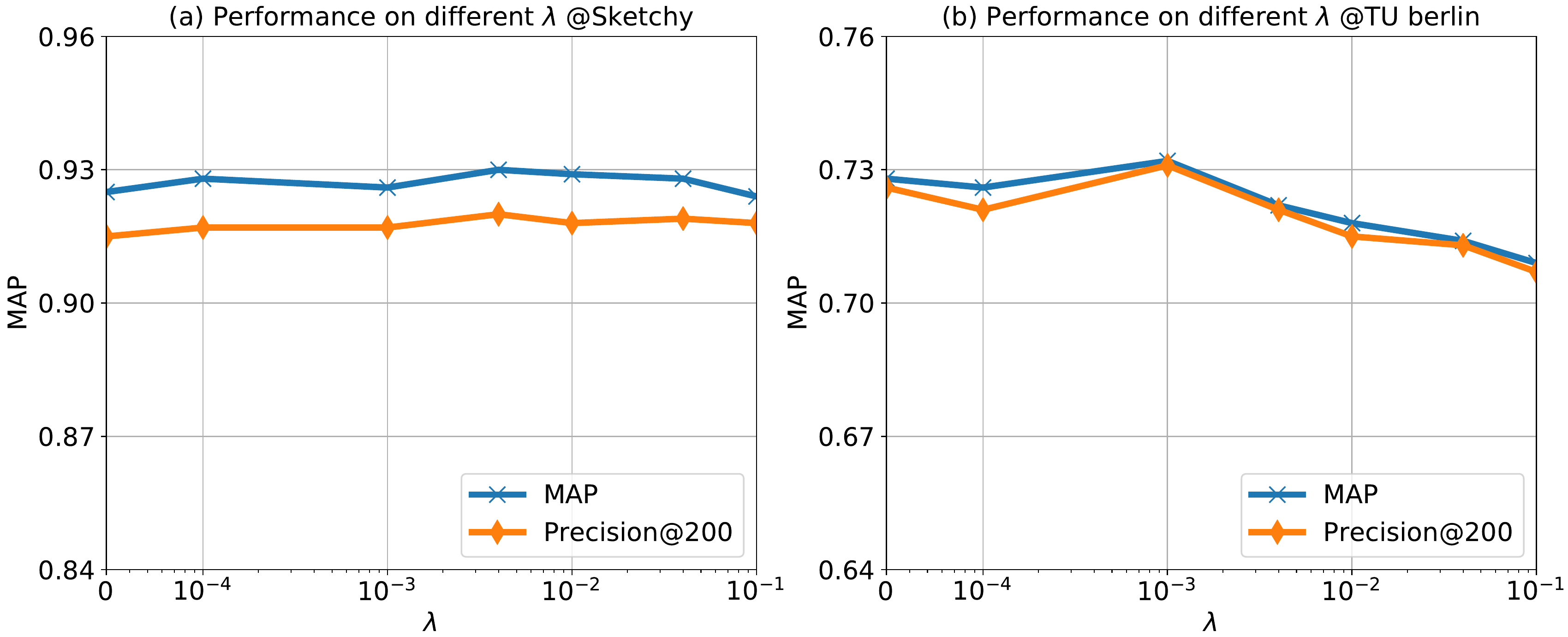}
    \caption{Comparative results in terms of MAP of different hyper-parameters on image-sketch dataset. The code length is 64.}
    \label{fig:params_sketch}
\end{figure}

\subsection{Sensitivity to Parameters}

In this section, the influence on different parameters of the proposed JCCH algorthm is evaluated. We use AlexNet for pre-training, and the code length is 64. We do not evalute the influence on $\alpha$ and $\mu$ as they have been discussed in~\cite{zhang2018semantic}.

\textbf{Influence on $\lambda$} Figure \ref{fig:params_sketch}(a)(b) and \ref{fig:params}(a1)(b1)(c1) shows the performance of different $\lambda$ on sketch-to-image and image-text retrieval tasks. It shows that a relatively small $\lambda$ is able to achive good MAP results and the performance is not sensitive to small $\lambda$. As discussed in~\cite{zhang2018semantic}, larger $\lambda$ makes $|\mathbf{h}_{i1} - \mathbf{c}_s|, |\mathbf{h}_{i2} - \mathbf{c}_s|$ go small to zero, making it hard to train or even collapsed.

\textbf{Influence on $\beta$} Figure \ref{fig:params}(a2)(b2)(c2) shows the performance on different $\beta$ on image-text retrieval tasks. It shows that a proper $\beta$ should be set for paired data to make the different modalities better aligned.

\section{Conclusions}

In this paper, we propose a novel and efficient cross-modal hashing algorithm named JCCH with just $O(n)$ compleixty. First, we proposed an improved Unary Upper Bound (UUB) for structured multilabel data in which the labels are highly unbalanced and correlated. The improved UUB is a more general and accurate bound for triplet ranking loss, and experiments convey that training with the improved UUB is able to achieve more efficient hashing performance. Second, we introduce the Corss-modal Unary Loss (CMUL) in which the improved UUB is introduced. The CMUL bridge the theoretical relationship between the cross-modal triplet loss and the $O(n)$ unary loss, and it is easy to be implemented as the model of each modality can be trained separately except that the semantic cluster centers are shared to make the hashcodes aligned. Third, we propose a novel cross-modal hashing algorithm called Joint Cluster Cross-modal Hashing (JCCH) in which the CMUL is introduced. The training procedure is expected to be efficient as the complexity is just $O(n)$. Experimental results on large-scale cross-modal datasets, including image-text datasets and sketch-image datasets, demonstrate that the training is more faster than other baselines, and the proposed method is superior over of comparable with the state-of-the-art cross-modal hashing algorithms.

\appendix
\section{The Brief Derivation of Improved Unary Upper Bound for Structured Multilabel Data}

In this section, we should prove the correctness of Eq. (\ref{eq:ot_mtuub}) and Algorithm \ref{alg:determine}. Considering Eq. (\ref{eq:trf}) and Eq. (\ref{eq:ml_triest}), we have the following bound for triplet ranking loss:
\begin{equation}
\begin{split}
\mathcal{L}_{ro} &\le \sum_{(i,j) \in S, (i,k) \notin S} \frac{1}{|Y_i \cap Y_j|} \sum_{s \in Y_i \cap Y_j} \frac{1}{|Y_k|} \sum_{t \in Y_k} \big[ g(|\mathbf{h}_i-\mathbf{c}_s|, |\mathbf{h}_i-\mathbf{c}_t|) \\
& \qquad \qquad \qquad \qquad + (|\mathbf{h}_j-\mathbf{c}_{s}|+|\mathbf{h}_k-\mathbf{c}_{t}|) \big] \\
\end{split}
\end{equation}

Compared with Eq. (\ref{eq:ot_mtuub_mid}), the coefficients $q_{ist} (s \neq t)$ can be computed as
\begin{equation}
\begin{split}
q_{ist} &= \sum_{j,k: (i,j) \in S, (i,k) \notin S} \frac{1}{|Y_i \cap Y_j|} \frac{1}{|Y_k|} 1_{s \in Y_i \cap Y_j, t \in Y_k} \\
&= \sum_{j: s\in Y_i \cap Y_j} \frac{1}{|Y_i \cap Y_j|} \sum_{k: t\in Y_k, (i,k) \notin S} \frac{1}{|Y_k|}
\end{split}
\end{equation}

And denote $u_{is} = u_{is}^{(1)} + u_{is}^{(2)}$, we first of all compute the $u_{js}^{(1)}$ by enumerating $|\mathbf{h}_j - \mathbf{c}_s|$:
\begin{equation}
\begin{split}
u_{js}^{(1)} &= \sum_{i:(i,j) \in S} \sum_{k: (i,k) \notin S} \frac{1}{|Y_i \cap Y_{j}|} \frac{1}{|Y_k|} \sum_{t \in Y_k} 1_{s \in Y_i \cap Y_{j}, t \in Y_k} \\
&= \sum_{i=1}^n \frac{1}{|Y_i \cap Y_j|} 1_{s \in Y_i \cap Y_j} \sum_{k: (i,k) \notin S} 1
\end{split}
\end{equation}
and then compute $u_{kt}^{(2)}$ by enumerating $|\mathbf{h}_k - \mathbf{c}_t|$:
\begin{equation}
\begin{split}
u_{kt}^{(2)} &= \sum_{i:(i,k) \notin S} \sum_{j: (i,j) \in S} \sum_{s \in Y_i \cap Y_j} \frac{1}{|Y_i \cap Y_j|} \frac{1}{|Y_k|} 1_{s \in Y_i \cap Y_j, t \in Y_k} \\
&= \sum_{i:(i,k) \notin S} \frac{1}{|Y_k|} \sum_{j: (i,j) \in S} 1
\end{split}
\end{equation}

The following notations are either from the original paper or from the Algorithm \ref{alg:determine}. We first of all consider that the anchor set $\mathbf{a}$ is the training set in the Algorithm \ref{alg:determine}. For $q_{ist}$, it is clear that $\frac{1}{|Y_i \cap Y_j|} 1_{s\in Y_i \cap Y_j} = \mathbf{L}_i^{s'}[j',s], (\mathbf{p}_i[j'] = j)$, thus we have $\sum_{j:s\in Y_i \cap Y_j} \frac{1}{|Y_i \cap Y_j|} = \mathrm{sum}(\mathbf{L}_i^{s'}, 0)[s]$. Similarly, $\frac{1}{|Y_k|} = \mathbf{Y}'[k,t]$ and $\sum_{k:t \in Y_k, (i,k) \notin S} \frac{1}{|Y_k|} = \mathrm{sum}(\mathbf{Y}'[\mathbf{n}_i], 0)[t]$. Thus we can get $q_{ist} = \mathrm{sum}(\mathbf{L}_i^{s'}, 0)[s] \cdot \mathrm{sum}(\mathbf{Y}'[\mathbf{n}_i], 0)[t]$, which is Line 11 in Algorithm \ref{alg:determine}.

For $u_{js}^{(1)}$, as $\frac{1}{|Y_i \cap Y_j|} 1_{s \in Y_i \cap Y_j} = \mathbf{L}_i^{s'}[j',s], (\mathbf{p}_i[j'] = j)$ and $\sum_{k:(i,k) \notin S} 1 = |\mathbf{n}_i|$, we have $u_{js}^{(1)} = \sum_{i=1}^n |\mathbf{n}_i| \mathbf{L}_i^{s'}[j',s] \cdot 1_{j \in \mathbf{p}_i}, (\mathbf{p}_i[j'] = j)$. It corresponds to Line 17 in Algorithm \ref{alg:determine}.

For $u_{kt}^{(2)}$, as $\frac{1}{|Y_k|} = \mathbf{Y}'[k,t]$ and $\sum_{j:(i,j) \in S} 1 = |\mathbf{p}_i|$, we have $u_{kt}^{(2)} = \sum_{i=1}^n |\mathbf{p}_i| \mathbf{Y}'[\mathbf{n}_i][k',t], (\mathbf{n}_i[k'] = k)$. It corresponds to Line 18 in Algorithm \ref{alg:determine}.

According the above derivations, the correctness of Algorithm \ref{alg:determine} is proved when the anchor set is the training set, and the complexity for determining the coefficients is reduced from $O(n^3)$ to $O(n^2)$. 

In practical applications, we need to sample a small anchor set from the training data, denote $\mathbf{a}, |\mathbf{a}| = l$, to reduce the complexity to $O(n)$. For computing $q_{ist}$, we just sample $j, k: (i,j) \in S, (i,k) \notin S$ from the anchor set and then $q_{ist} = (\frac{n}{l})^2 \mathrm{sum}(\mathbf{L}_i^{s'}, 0)[s] \cdot \mathrm{sum}(\mathbf{Y}'[\mathbf{n}_i], 0)[t]$. For computing $u_{js}^{(1)}$ and $u_{kt}^{(2)}$, we just sample $i$ from the anchor set and then $u_{js}^{(1)} = \frac{n}{l} \sum_{i \in \mathbf{a}} |\mathbf{n}_i| \mathbf{L}_i^{s'}[j',s] \cdot 1_{j \in \mathbf{p}_i}, (\mathbf{p}_i[j'] = j), u_{kt}^{(2)} = \frac{n}{l} \sum_{i \in \mathbf{a}} |\mathbf{p}_i| \mathbf{Y}'[\mathbf{n}_i][k',t], (\mathbf{n}_i[k'] = k)$. Note that the above derivations correspond to  Line 7-10,14-16 in Algorithm \ref{alg:determine}. The term $(\frac{n}{l})^2$ and $\frac{n}{l}$ hold in the fact that we just sample $(\frac{l}{n})^2$ of all triplets for computing $q_{ist}$ and sample $\frac{l}{n}$ of all triplets for computing $u_{is}$. To conclude, the coefficients can be arrived in $O(n)$ time with Line 11,17,18,20 in Algorithm \ref{alg:determine}, thus the Algorithm \ref{alg:determine} is correct. 

